\pdfoutput=1
\documentclass{elsarticle}
\usepackage{subfigure}
\usepackage{amsmath}    
\usepackage{bm}
\usepackage{graphicx}   
\usepackage{array}
\usepackage{subfigure}  

\renewcommand\vec[1]{\bm{\mathrm{#1}}}

\begin{document}

\title{\textsc{BoltzWann}: A code for the evaluation of thermoelectric and electronic transport properties with a maximally-localized Wannier functions basis}

\author{Giovanni Pizzi\fnref{fn1}}
 \address{Theory and Simulation of Materials, \'Ecole Polytechnique F\'ed\'erale de Lausanne, 1015 Lausanne, Switzerland}
 \ead{giovanni.pizzi@epfl.ch}   
\author{Dmitri Volja\fnref{fn1}}
 \address{Department of Materials Science and Engineering,  Massachusetts Institute
of Technology,  Cambridge,
Massachusetts 02139, USA}
\author{Boris Kozinsky}
\address{Research and Technology Center, Robert Bosch LLC, Cambridge, Massachusetts 02139, USA}
\author{Marco Fornari}
\address{Department of Physics, Central Michigan University, Mount Pleasant, Michigan 48859, USA}
\author{Nicola Marzari}
\address{Theory and Simulation of Materials, \'Ecole Polytechnique F\'ed\'erale de Lausanne, 1015 Lausanne, Switzerland}

\fntext[fn1]{These two authors equally contributed to the work.}

\begin{keyword}
Maximally-localized Wannier functions \sep Wannier90 \sep band interpolation \sep band velocities \sep thermoelectric properties
\end{keyword}

\begin{abstract}
We present a new code to evaluate thermoelectric and electronic transport properties of extended systems with a maximally-localized Wannier function basis set. The semiclassical Boltzmann transport equations for the homogeneous infinite system are solved in the constant relaxation-time approximation and band energies and band derivatives are obtained via Wannier interpolations. Thanks to the exponential localization of the Wannier functions obtained, very high accuracy in the Brillouin zone integrals can be achieved with very moderate computational costs. Moreover, the analytical expression for the band derivatives in the Wannier basis resolves any issues that may occur when evaluating derivatives near band crossings. The code is tested on binary and ternary skutterudites CoSb$_3$ and CoGe$_{3/2}$S$_{3/2}$.
 \end{abstract}

\maketitle

\section{Introduction}
Electrical and heat conductivities are fundamental properties characterizing a crystal and their evaluation allows to determine whether a given material is suitable or not for a specific application.
For instance, a piezoelectric material should be an electrical insulator to avoid current leakages when an electric field is applied~\cite{Armiento:2011}, while other applications as electronic or thermoelectric devices may require a conductor. 
Recently, nanostructured materials have also attracted attention in the literature. The reduced size of such systems can change significantly their heat transport properties or improve their performance in specific applications. This happens for example in sintered powder nanocomposites, which can display a thermoelectric efficiency better than their bulk counterparts~\cite{Poudel:2008,Xie:2009}.

Advances in simulation techniques and the increase of
computational power allow nowadays to calculate and predict such properties in
inexpensive ways and compare with experiments.
Moreover, to study and design new materials, it becomes ever more important to be able to predict material properties even before the material is experimentally synthesized, providing in this way guidance and search criteria (see for instance Ref.~\cite{Wang:2011}).
For this aim, it is convenient to use theoretical techniques which are general enough to take into account broad classes of materials, as it is the case for plane-wave methods based on density-functional theory that we will use in this paper~\cite{Payne:1992,Giannozzi:2009}.

Among the transport properties of crystals, thermal and electrical transport is currently being intensively studied in the literature, due to the interest in finding more efficient thermoelectric materials to be used for energy harvesting and waste heat recovery~\cite{Wang:2011,Mahan:1996,Snyder:2008,Vineis:2010,Li:2010}.
To assess the performance of a thermoelectric material, one can use the figure of merit $ZT$, defined by
\begin{equation}
\label{zt}
ZT=\frac{\sigma S^2T}{\kappa},
\end{equation}
where $\sigma$ is the electrical conductivity, $S$ is the Seebeck coefficient, $\kappa$ is the thermal conductivity (electronic+ionic) and $T$ is the temperature (see e.g.~\cite{Sharp:1995}).
Since a large $ZT$ coefficient indicates that the material could be a good thermoelectric, one should therefore try to maximize the Seebeck coefficient $S$ and the electrical conductivity $\sigma$ and at the same time reduce the thermal conductivity $\kappa$. These properties, however, are coupled and a balance between them is required in order to achieve a high figure of merit.
Currently, the best-performing commercially available bulk thermoelectric materials have a $ZT$ slightly above 1 at room temperature or above~\cite{Vineis:2010,Venkatasubramanian:2001}, while the highest ZT for a laboratory material engineered at the nano- and microscopic scale is 2.2~\cite{Biswas:2012}.

From the theoretical point of view, while for the calculation of the density of states (DOS) one needs only the band energy dispersion $E_{n,\vec k}$ over the Brillouin zone, for the evaluation of electrical and thermal transport properties also band velocities are required (see Sec.~\ref{sec:semiclassical-transport}):
\begin{equation}
v_{i}(n,\vec{k})=\frac{1}{\hbar}\frac{\partial E_{n,\vec{k}}}{\partial k_{i}}. \label{eqn:bandvelocity}
\end{equation}

A typical approach for the calculation of $v_{i}(n,\vec{k})$ is to evaluate the band structure on a dense $\vec k$ grid in the Brillouin zone in reciprocal space, and to use a finite-difference method to evaluate its derivatives. To achieve convergence on transport properties, however, a very dense $\vec k$ grid is required due to the steepness of the Fermi--Dirac distribution near the Fermi energy. Since with first-principles methods the solution of the eigenvalue problem is costly (in general it scales as $O(N^3)$, where $N$ is the system size, and is directly proportional to the number of $\vec k$ points in the Brillouin zone), some form of interpolation of the bands is extremely beneficial~\cite{Pickett:1988,Uehara:2000,Madsen:2006,Chaput:2005} in order to complete the calculation within reasonable computation times, even more so if more accurate and costly electronic-structure approaches (such as many-body perturbation theory) are used~\cite{Hamann:2009}.
This finite-difference procedure for the calculation of the band velocities, however, may be a source of error if the interpolation provides an incorrect band ordering near crossings.
Furthermore, only bands in an energy region of width $KT$ around the chemical potential $\mu$ are relevant; thus, in semiconductor systems we only need a detailed description of the bands near the band gap. It is then highly desirable to use models in which only this relevant energy window  is considered.

For all the above reasons, we present in this work our \textsc{BoltzWann} code to evaluate electron transport properties in a semiclassical formalism using a maximally-localized Wannier function (MLWF) basis~\cite{Marzari:1997,Souza:2001} to interpolate first-principles plane-wave (PW) results. 
With this approach, we can accurately evaluate transport quantities by solving the eigenvalue problem on a coarse $\vec k$-point reciprocal-space grid. 
Once MLWFs have been obtained from this coarse $\vec k$ grid, they can be used as a tight-binding basis to interpolate the band structure on a denser $\vec k$ grid at a significantly smaller computational cost~\cite{Souza:2001,Lee:2005,Yates:2007}.
Furthermore, thanks to the strong exponential localization of the WFs (see \cite{Brouder:2007} and Secs. II.G and VI.A of~\cite{Marzari:2012}), the results can match with excellent accuracy the full plane-wave calculations~\cite{Souza:2001}.
Finally, in the Wannier representation, the band derivatives can be evaluated analytically at an arbitrary $\vec k$ point without the need to rely on finite-difference methods (see Sec.~\ref{sec:velocity-interp}), producing numerically stable results even at band crossings and near weak avoided crossings~\cite{Yates:2007}.

The paper is organized as follows. In Sec.~\ref{sec:semiclassical-transport} we summarize the most relevant results of the semiclassical Boltzmann transport theory that are used in the \textsc{BoltzWann} code. Then, in Sec.~\ref{sec:wf-basis}, we briefly introduce some fundamentals in the definitions and use of Wannier functions, and in particular on how they can be used to interpolate band velocities at a generic $\vec k$ point. 
In Sec.~\ref{sec:code-details} the \textsc{BoltzWann} code is presented and the relevant input parameters are described. 
Finally, as a verification, in Sec.~\ref{sec:application} we use \textsc{BoltzWann} to evaluate the thermal properties of two skutterudite systems (CoSb$_3$ and CoGe$_{3/2}$S$_{3/2}$). The results are compared with those obtained with the \textsc{BoltzTraP} code~\cite{Madsen:2006}, which uses a different scheme to interpolate the bands. A comparison with the experimental Seebeck coefficient of CoSb$_3$~\cite{Caillat:1996,Volja:2012} is also provided.

\section{Semiclassical transport theory}
\label{sec:semiclassical-transport}
The basic transport equations for the current density $\vec J$ and the electronic heat current ${\vec J_Q}$ for a system in the
presence of an electric field $\vec E$ and a temperature gradient $\vec \nabla T$ are given by~\cite{Madsen:2006,Ziman:1972,Grosso:2000}: 
\begin{align}
 \label{currentdensity}
{\vec {J}}& ={\bm {\sigma}}({ \vec{E}}- {\bf S}\vec\nabla T),\\
 \label{heatcurrent}
{\vec J_Q}&= T{\bm \sigma}{\bf S}{\vec {E}} - {\bm  K}\vec \nabla T,
\end{align}
where  ${\bm \sigma}$, $\bf S$ and $\bf K$ are rank-two tensors which reduce to scalars for isotropic media; ${\bm \sigma}$ is the electrical conductivity, $\bf S$ is the  Seebeck coefficient, and the thermal conductivity $\bm \kappa$ is defined as (minus) the heat current per unit of temperature gradient in open-circuit conditions (i.e., $\vec J=0$) and thus it is given by
$$
\bm \kappa = \bm K - T \bm \sigma \bm S^2.
$$
We stress that the heat current $\vec J_Q$ discussed here contains only the electronic contribution to the thermal transport; to take into account also the ionic contribution, phonon--phonon scattering processes must be taken into account~\cite{Srivastava:1990,Garg:2011}. Note also that, especially in semiconductors and insulators, the lattice contribution to the thermal conductivity is the dominating one.

We derive the expressions for the tensors $\bm \sigma$, $\bm S$ and $\bm K$ in the framework of the semiclassical transport theory using the Boltzmann transport equation. 
A convenient general way of describing the collisional term in the Boltzmann equation is to define a lifetime $\tau_{n\vec{k}}$ for an electron on band $n$ at wavevector $\vec{k}$. 
Then, we obtain the following expressions for the transport tensors as a function of the chemical potential $\mu$ (that depends on the doping of the system in a semiconductor) and of the temperature $T$~\cite{Ziman:1972,Scheidemantel:2003,Mahan:2006}:
\begin{align}\label{eq:eq-sigma} [\bm{\sigma}]_{ij}(\mu,T)&=
  e^{2}\int_{-\infty}^{+\infty}dE\left(-\frac{\partial
      f(E,\mu,T)}{\partial E}\right)\Sigma_{ij}(E),\\
\label{eq:eq-seebeck}  [\bm{\sigma}\bm{S}]{}_{ij}(\mu,T)&=
  \frac{e}{T}\int_{-\infty}^{+\infty}dE\left(-\frac{\partial
      f(E,\mu,T)}{\partial E}\right)(E-\mu)\Sigma_{ij}(E),\\
\label{eq:eq-thermal}  [\bm{K}]{}_{ij}(\mu,T)&=
  \frac{1}{T}\int_{-\infty}^{+\infty}dE\left(-\frac{\partial
      f(E,\mu,T)}{\partial E}\right)(E-\mu)^{2}\Sigma_{ij}(E).
\end{align}
Here, $i$ and $j$ are Cartesian indices, $\bm \sigma \bm S$ denotes the matrix product of the two tensors, and $\partial f/\partial E$ is the derivative of the Fermi--Dirac distribution function with respect to the energy. Moreover, we have defined the above quantities in terms of the transport distribution function (TDF) $\Sigma_{ij}(E)$, defined as
\begin{equation}
\Sigma_{ij}(E)=\frac{1}{V}\sum_{n,\vec{k}}v_{i}(n,\vec{k})v_{j}(n,\vec{k})\tau_{n\vec{k}}\delta(E-E_{n,\vec k}),
 \label{eq:sigma-ij}
\end{equation}
where the summation is over all bands $n$ and over all the Brillouin zone (BZ), $E_{n,\vec k}$ is the energy for band $n$ at $\vec k$ and $v_i$ is the $i-$th component of the band velocity at $(n,\vec k)$ given in Eq.~\eqref{eqn:bandvelocity}.

The most simple approximation that we adopt for the results shown in Sec.~\ref{sec:application} is to assume that the lifetime $\tau_{n\vec{k}}$ is independent both of $n$
  and $\vec{k}$, and to choose the value $\tau=\tau_{n\vec{k}}$ by fitting the experimental values for, e.g., the experimental electrical conductivity at a given temperature.
This constant relaxation-time approximation is often adopted in first-principles calculations~\cite{Madsen:2006a,Scheidemantel:2003}, and despite its simplicity it often turns out to be a good approximation for bulk materials, even in the case of anisotropic systems~\cite{Allen:1988}. 
This approximation can be relaxed if one is provided with a model for $\tau$. For instance, one can take into account an energy dependence for the relaxation time~\cite{Allen:1988} or calculate $\tau_{n,\vec k}$ from first-principles considering explicitly the electron-phonon scattering mechanisms~\cite{Mauri:1996,Bonini:2007,Giustino:2007,Noffsinger:2010}.
For sintered powders, instead, a different model is more physically motivated, i.e., the assumption of a constant mean free path, which is taken to be of the order of the grain size~\cite{Wang:2011}.
In any case, the {\textsc{BoltzWann}} code presented here can be simply adapted to take into account any advanced model for lifetimes by using a $n-$ and $\vec k-$dependent $\tau_{n,\vec k}$ in Eq.~\eqref{eq:sigma-ij}.

\section{Wannier functions basis}
\label{sec:wf-basis}
We outline in this Section the method to obtain MLWFs; further details can be found in Refs.~\cite{Marzari:1997,Souza:2001,Mostofi:2008,Marzari:2012}.

According to Bloch theorem, 
$\psi_{n\vec k}(\vec r)=u_{n\vec k}(\vec r)e^{i\vec k\cdot\vec r}$
are the wavefunctions of the crystal labeled by a band index $n$ and a crystal momentum $\vec k$, where the Bloch part $u_{n\vec k}(\vec r)$ has the periodicity of the crystal.
If we are interested only in the valence band of a semiconductor (for instance, to simulate only $p-$doped systems), and this is composed by $N$ bands, we can choose to construct  from this $N-$dimensional manifold $N$ Wannier functions in each unit cell with the transformation
\begin{equation}
  \label{eq:wannier-1}
  |\vec R n\rangle = \frac V{(2\pi)^3}\int_{\text{BZ}} d\vec k \sum_{m=1}^N e^{-i \vec k\cdot \vec R}\mathcal U_{mn}(\vec k)|\psi_{n,\vec k}\rangle,
\end{equation}
where $\mathcal U_{mn}(\vec k)$ is a unitary matrix, chosen so as to minimize the localization functional $\Omega$ defined as~\cite{Marzari:1997}
\begin{eqnarray}
 \label{omega}
\Omega=\sum_{n=1}^N (\langle 0n|r^2|0n\rangle-\langle 0n|{\vec r}|0n\rangle^2).
\end{eqnarray}

If we need instead to deal with ``entangled'' bands, as it is e.g. the case for the 
conduction-band manifold, the method must be extended according to the prescriptions in Ref.~\cite{Souza:2001} by performing a preliminary step to separate or ``disentangle'' a $N-$dimensional manifold from which the Wannier functions can then be constructed.
To this aim, two energy windows are defined.
First, an outer energy window (with $E<E_{\text{max}}$) is considered, and all original Bloch states within this window are considered to construct WFs (note that the number $N_{\vec k}$ of Bloch states in this window 
may vary from one $\vec k$ point to another).
Then, an inner ``frozen'' window may be additionally defined (with $E<E_{\text{froz}}$), where all Bloch states are forced to be preserved. 
At this point, one has to map for each $\bf k$ (via a set of $N_{\vec k}\times N$ rectangular matrices) the $N_{\bf k}$ states onto an equal or smaller number of $N$ orthonormal Bloch-like states,
from which the Wannier functions can be extracted by minimizing $\Omega$ as described above.
This mapping is obtained by enforcing a requirement of optimal smoothness across the Brillouin zone of the ``disentangled'' subspace (see Ref.~\cite{Souza:2001}).
In this way, one obtains at each $\vec k$ point 
a new set of Bloch functions, $u^{(W)}_n(\vec k)$, that are those used to construct MLWFs by means of the transform in Eq.~\eqref{eq:wannier-1}  (the superscript $(W)$ indicates that these are the Bloch states used to construct MLWFs, and define the ``Wannier'' gauge). 

We can now project the Hamiltonian operator $\hat H$ onto the $N-$dimensional subspace of the WFs, obtaining a $N\times N$ matrix:
\begin{eqnarray}
H_{mn}^{(W)}({\bf k})=\langle u^{(W)}_{n}({\bf k})| \hat H({\bf k})| u^{(W)}_{m}({\bf k}) \rangle,
\label{hwan}
\end{eqnarray}
where $\hat H(\vec k) = e^{-i\vec k\cdot \vec r}\hat He^{i\vec k\cdot \vec r}$ and
the Hamiltonian $\hat H$ on the original first-principles basis $|u_{n\vec k}\rangle$ is a diagonal  $N_{\bf k}\times N_{\bf k}$ matrix: $H_{nm}({\bf k})=E_{n{\bf k}}\delta_{nm}$.

Substituting the definition of the $| u_m^{(W)}\rangle$ in terms of the $|u_n\rangle$ in the above equation, one can easily show that $H_{mn}^{(W)}({\bf k})$ is simply the Fourier transform of the matrix elements of $\hat H$ between Wannier functions:
\begin{equation}
  \label{eq:HW-is-FT-of-Wannier-matrixelements}
H_{nm}^{(W)}(\vec k) = \sum_{\vec R} e^{i\vec k\cdot \vec R} \langle n\vec 0|\hat H|m\vec R\rangle.
\end{equation}
However, since the WFs are not eigenstates of the energy, $H^{(W)}$ is not diagonal; we can thus change gauge and finally obtain a diagonal $H^{(H)}$ matrix by means of a $N\times N$ unitary matrix {\it U}
\begin{eqnarray}
 U^\dagger({\bf k})H^{(W)}({\bf k})U({\bf k})=H^{(H)}({\bf k} ),
\label{umat}
\end{eqnarray}
where the superscript $(H)$ denotes the Hamiltonian gauge spanned by the functions
\begin{eqnarray}
 \label{blochh}
|u^{(H)}_{n}({\bf k})\rangle=\sum_m^{N} |u^{(W)}_{m} ({\bf k}) \rangle U_{mn}({\bf k}).
\end{eqnarray}
Note that, inside the inner energy window, the eigenvalues of $H^{(H)}$ coincide with the original eigenvalues $E_{n\vec k}$, and the corresponding functions $|u^{(H)}_{n}({\bf k})\rangle $ in the Hamiltonian representation match with the original Bloch states $|u_n({\bf k})\rangle$.

\subsection{Interpolation of the velocity operator at an arbitrary $\vec k$ point}
\label{sec:velocity-interp}
The evaluation of the velocity operator for an arbitrary $\vec k$ point is in general nontrivial~\cite{Read:1991,Kageshima:1997,Hybertsen:1987}. We focus here in particular on the quantity of our interest, i.e. the band velocity of Eq.~\eqref{eqn:bandvelocity}, for the case of non-degenerate bands:
\begin{equation}
  \label{eq:bandvelocity-vs-matrixelems}
  v_{n\vec k,\alpha} \equiv \frac 1 \hbar \frac{\partial E_{n\vec k}}{\partial k_\alpha}=
\frac 1 \hbar \frac{\partial }{\partial k_\alpha}\left[
  \langle u_{n\vec k}^{(H)} | H^{(H)}|u_{n\vec k}^{(H)}\rangle
\right],
\end{equation}
where in the last step we have used the fact that $H^{(H)}$ is diagonal and that, inside the inner window, the eigenvalues of $H^{(H)}$ coincide with those of the first-principles $H$ matrix.

In order to obtain an analytic expression for the band velocities, we need to obtain an expression for $v_{n\vec k,\alpha}$ as a function of $\partial_\alpha H^{(W)}$ (we use the shortcut notation $\partial_\alpha$ to indicate the partial derivative $\partial/\partial k_\alpha$), since $\partial_\alpha H^{(W)}$ can be obtained in a simple way from the WF matrix elements. In fact, taking the derivative of Eq.~\eqref{eq:HW-is-FT-of-Wannier-matrixelements}, we simply get
\begin{equation}
  \label{eq:HW-derivative}
  \partial_\alpha H_{nm}^{(W)} = \sum_{\vec R} e^{i\vec k\cdot \vec R} iR_\alpha \langle n\vec 0|\hat H| m\vec R\rangle.
\end{equation}
Let us start by taking explicitly the derivative in~\eqref{eq:bandvelocity-vs-matrixelems}:
\begin{align}
  \label{eq:bandvelocity-vs-matrixelems-2}
  \hbar v_{n\vec k,\alpha} =&
  E_{n\vec k} \langle \partial_\alpha u_{n\vec k}^{(H)} | u_{n\vec k}^{(H)}\rangle + 
  E_{n\vec k} \langle u_{n\vec k}^{(H)} | \partial_\alpha  u_{n\vec k}^{(H)}\rangle +\\
&  + \langle u_{n\vec k}^{(H)} |\partial_\alpha H^{(H)}| u_{n\vec k}^{(H)}\rangle.
\end{align}
The first two terms on the right-hand cancel out, since $\langle u_{n\vec k}^{(H)} | u_{n\vec k}^{(H)}\rangle =1 $ and thus 
\begin{equation*}
0=\partial_\alpha \langle u_{n\vec k}^{(H)} | u_{n\vec k}^{(H)}\rangle \quad \Rightarrow \quad
\langle \partial_\alpha u_{n\vec k}^{(H)} | u_{n\vec k}^{(H)}\rangle = - \langle u_{n\vec k}^{(H)} | \partial_\alpha u_{n\vec k}^{(H)}\rangle.
\end{equation*}
Moreover, using Eq.~\eqref{umat} and the unitarity of $U$, we can rewrite
\begin{equation}
  \label{eq:HH-function-HW}
  \partial_\alpha H^{(H)} = D_\alpha^\dagger H^{(H)} + H^{(H)}D_\alpha + U^\dagger (\partial_\alpha H^{(W)}) U,
\end{equation}
where following Ref.~\cite{Yates:2007} we have defined  $D_\alpha = (U^\dagger \partial_\alpha U)$.
Since $U$ is unitary, $D_\alpha$ is antihermitian; moreover, being $H^{(H)}$ diagonal, the diagonal elements of $D_\alpha^\dagger H^{(H)} + H^{(H)}D_\alpha$ are zero. Thus, finally, substituting Eq.~\eqref{eq:HH-function-HW} in Eq.~\eqref{eq:bandvelocity-vs-matrixelems-2} we get  the required expression for the band velocities
\begin{equation}
  \label{eq:bandvelocity-final}
  v_{n\vec k,\alpha} = \frac 1 \hbar \left\langle u_{n\vec k}^{(H)} \left|U^\dagger \left(\partial_\alpha H^{(W)}\right) U \right|u_{n\vec k}^{(H)}\right\rangle,
\end{equation}
i.e., they are the diagonal elements of the matrix product $[U^\dagger (\partial_\alpha H^{(W)}) U]$.

We emphasize here that for the band velocities (i.e., first derivatives) the result is simple thanks to the cancellation of the terms described above. However, if one is also interested in second derivatives (e.g., for the calculation of the effective masses), non-diagonal terms do also contribute and the proper treatment of such terms needs to be coded. For a detailed discussion, which takes into account also the case of band degeneracies, we refer to Ref.~\cite{Yates:2007}.

\section{Details of the code}
\label{sec:code-details}
The \textsc{BoltzWann} code has been developed as a module of the \textsc{Wannier90} package~\cite{Mostofi:2008}, available at {\sc www.wannier.org}.
A list of all input parameters that control  \textsc{BoltzWann} is reported in Table~\ref{tab:input}. They must be included in the same \texttt{seedname.win} input file of \textsc{Wannier90}, and they are all prefixed by \texttt{boltz} to distinguish them from other variables of the code.

\begin{table}
\begin{center}
\begin{tabular}{l>{\small}p{6cm}}
\emph{Input flag} & \emph{Description}\\
\hline
{\tt boltzwann}   & Flag to activate the \textsc{BoltzWann} module \\
{\tt boltz\_kmesh} & Interpolation $\vec k$-mesh (one or three integers)\\ 
{\tt boltz\_kmesh\_spacing} & Minimum spacing between $\vec k$ points in \AA$^{-1}$\\
{\tt boltz\_relax\_time} & Relaxation time in fs\\
{\tt boltz\_mu\_min} & Minimum value of $\mu$ in eV\\
{\tt boltz\_mu\_max} & Maximum value of $\mu$ in eV\\
{\tt boltz\_mu\_step} & Step for $\mu$ in eV\\
{\tt boltz\_temp\_min} & Minimum value of the temperature $T$ in K \\
{\tt boltz\_temp\_max} & Maximum value of the temperature $T$ in K \\
{\tt boltz\_temp\_step} & Step for $T$ in K \\
{\tt boltz\_tdf\_energy\_step} & Energy step for the TDF (eV) \\
{\tt boltz\_tdf\_smr\_fixed\_en\_width} & Energy smearing for the TDF (eV) \\
{\tt boltz\_tdf\_smr\_type} & Smearing type for the TDF \\
{\tt boltz\_calc\_also\_dos} & Flag to calculate also the DOS while calculating the TDF\\
{\tt boltz\_dos\_energy\_min} & Minimum energy value for the DOS in eV \\
{\tt boltz\_dos\_energy\_max} & Maximum energy value for the DOS in eV \\
{\tt boltz\_dos\_energy\_step} & Step for the DOS in eV\\
{\tt boltz\_dos\_smr\_type} & Smearing type for the DOS \\
{\tt boltz\_dos\_adpt\_smr} & Flag to use adaptive smearing for the DOS \\
{\tt boltz\_dos\_adpt\_smr\_fac} & Factor for the adaptive smearing \\
{\tt boltz\_dos\_adpt\_smr\_max} & Maximum allowed value for the adaptive energy smearing (eV) \\
{\tt boltz\_dos\_smr\_fixed\_en\_width} & Energy smearing for the DOS (eV) \\
{\tt boltz\_bandshift} & Flag to add a rigid bandshift to the conduction bands\\
{\tt boltz\_bandshift\_firstband} & Index of the first band to shift\\
{\tt boltz\_bandshift\_energyshift} & Energy shift of the conduction bands (eV)\\
\hline
\end{tabular}
\caption{Input keywords controlling the \textsc{BoltzWann} module. These flags are inserted in the \texttt{seedname.win} file together with the other input flags of \textsc{Wannier90}.}
\label{tab:input}
\end{center}
\end{table}

To execute \textsc{BoltzWann}, one first runs the standard \textsc{Wannier90} calculation using the {\tt wannier90.x} executable to obtain the maximally-localized Wannier functions. Once the MLWFs are obtained, the execution of the {\tt postw90.x} executable activates all post-processing modules that use the Wannier functions as an input (more detailed information can be found in the user manual and in the tutorial distributed with the code).
In particular, setting {\tt boltzwann=.true.} in the input file activates the \textsc{BoltzWann} module. As a first step, the code calculates the TDF function on a $n_1\times n_2\times n_3$ Monkhorst-Pack~\cite{Monkhorst:1976} $\vec k$ mesh (containing $\Gamma$), where $n_1$, $n_2$ and $n_3$ are defined by means of the {\tt boltz\_kmesh} input parameter. Alternatively, a minimum spacing between neighboring $\vec k$ points can be specified via the {\tt boltz\_kmesh\_spacing} input flag, and then the code finds the integers $n_1$, $n_2$ and $n_3$ that satisfy this requirement. The TDF $\bm \Sigma(E_i)$ is evaluated on a suitable energy grid $E_i$, with energy step $E_{i+1}-E_i$ defined by {\tt boltz\_tdf\_energy\_step}.
The time-consuming operation in the whole calculation is the interpolation
of the band structure on the dense $n_1\times n_2\times n_3$ $\vec k$ mesh. Thus, the preliminary evaluation of the 
TDF is convenient, because the interpolation has to be performed only once. Then, when the TDF is known, the transport properties can be calculated inexpensively using Eqs.~\eqref{eq:eq-sigma}, \eqref{eq:eq-seebeck} and \eqref{eq:eq-thermal} on an arbitrary set of $(\mu, T)$ pairs. For instance, for the calculations of Sec.~\ref{sec:application}, the time taken in this last part is one or two orders of magnitude smaller than the one required for the TDF evaluation.

For the same reason, we provide a flag ({\tt boltz\_calc\_also\_dos}) to calculate the density of states (DOS) using \textsc{BoltzWann}. In this way, we can use the same $\vec k$ grid to calculate also the DOS, hence performing the band interpolation only once when the DOS is also needed (for instance, for the comparison with an experimental sample with fixed doping, as discussed at the end of Sec.~\ref{sec:results-and-verification}).
Different kinds of smearing can be chosen for the DOS through the {\tt boltz\_dos\_smr\_type} flag: Gaussian, Marzari--Vanderbilt~\cite{Marzari:1999} and Methfessel--Paxton~\cite{Methfessel:1989}. Moreover, beside the standard fixed-width smearing, one can alternatively use the adaptive smearing introduced in Ref.~\cite{Yates:2007} (flag {\tt boltz\_dos\_adpt\_smr}) that exploits the knowledge of  band derivatives to obtain a $\vec k-$dependent smearing.
We also allow for a smearing of the TDF function by means of the flags {\tt boltz\_tdf\_smr\_fixed\_en\_width} and {\tt boltz\_tdf\_smr\_type}. This is intended to smooth the transport properties at low temperatures. However, smearing can introduce spurious effects and its use should be in general avoided (in fact, no smearing of the TDF is introduced by default).
Furthermore, if {\tt boltz\_bandshift=.true.}, a scissor operator can be applied to rigidly shift the conduction bands by a fixed energy chosen through the flag {\tt boltz\_bandshift\_energyshift}, in order to correct for the approximate band\-gaps provided by DFT. This shift is applied to all bands with index larger than {\tt boltz\_bandshift\_firstband}. This can be useful when comparing simulations with experiments, if one wants to take into account also the effect of the minority intrinsic carriers on the total number of free carriers.

Finally, the code is fully parallel. In particular, for the calculation of the TDF function, the $\vec k$ points of the mesh are distributed across processors. Each processor sums locally the contributions to the TDF from all the $\vec k$ points assigned to it, and only at the end the contributions from the different processors are summed together. The final TDF function is then distributed to all processors. For the second part of the calculation, the $(\mu,T)$ pairs are now distributed, and each processor calculates the $\bm \sigma$, $\bf S$ and $\bf K$ tensors on its subset of $(\mu,T)$ pairs.

\section{Applications and verification}
\label{sec:application}
In this section we present an application of the code to the binary skutterudite CoSb$_3$, and the promising ternary skutterudite CoGe$_{3/2}$S$_{3/2}$~\cite{Vaqueiro:2008,Volja:2012}.

We first briefly discuss the crystal structure used for the simulation of these two systems. The space group symmetry of CoSb$_3$ is Im$\overline 3$. The unit cell is body-centered cubic with 4 Co atoms and 12 Sb atoms, and lattice parameter 16.95~bohr. The unit cell contains four Co atoms with Wyckoff letter $8c$ and Wyckoff position $(0.25,0.25,0.25)$, and twelve Sb atoms with Wyckoff letter $24g$ and position $(0.0,0.3336,0.1590)$.   
The space group of CoGe$_{3/2}$S$_{3/2}$ is instead R$\overline 3$, with 8 Co atoms,
12 Ge atoms and 12 S atoms in the unit cell.
The unit cell is rhombohedral with lattice parameters $a=14.866$~bohr and $\cos\gamma=0.001855$.
The Wyckoff positions of the atoms are: Co atoms at $(2c)(0.259,0.259,0.259)$ and $(6f)(0.258,0.762,0.754)$; Ge atoms at $(6f)(0.999,0.335,0.151)$ and $(6f)$\linebreak[0]$(0.5,0.835,0.349)$; S atoms at $(6f)(0.,0.343,0.849)$ and $(6f)(0.502,0.844,0.650)$.

\subsection{First-principles computational details}
Calculations of the Bloch states $|u_{n\vec k}\rangle$ are performed using density-functional theory~\cite{Hohenberg:1964,Kohn:1965} in the LDA approximation~\cite{Ceperley:1980,Perdew:1981}.
We use ultrasoft pseudopotentials for Co and S atoms, and norm-conserving pseudopotentials for Sb and Ge, publicly available at www.quantum-espresso.org~\cite{Martin:2004,Vanderbilt:1990,Giannozzi:2009}. A plane-wave basis is adopted for the expansion of the valence electron wavefunctions and the charge density, with kinetic-energy cutoffs of 30~Ry and 240~Ry respectively.
Calculations have been performed using the \texttt{pw.x} code of the \textsc{Quantum ESPRESSO} distribution~\cite{Giannozzi:2009}.
For ground state calculations we use a $6 \times 6 \times 6$ Monkhorst-Pack $\vec k$-point mesh for CoSb$_3$ and $4\times 4\times 4$ $\vec k$-mesh for CoGe$_{3/2}$S$_{3/2}$. 
From the ground state density, the Bloch states and the overlaps for the calculation of the MLWFs have been computed on a $4\times 4\times 4$ mesh for both systems. 

\subsection{Wannier functions basis in CoSb$_3$}

\begin{figure}[tbp]
\includegraphics[width=12cm]{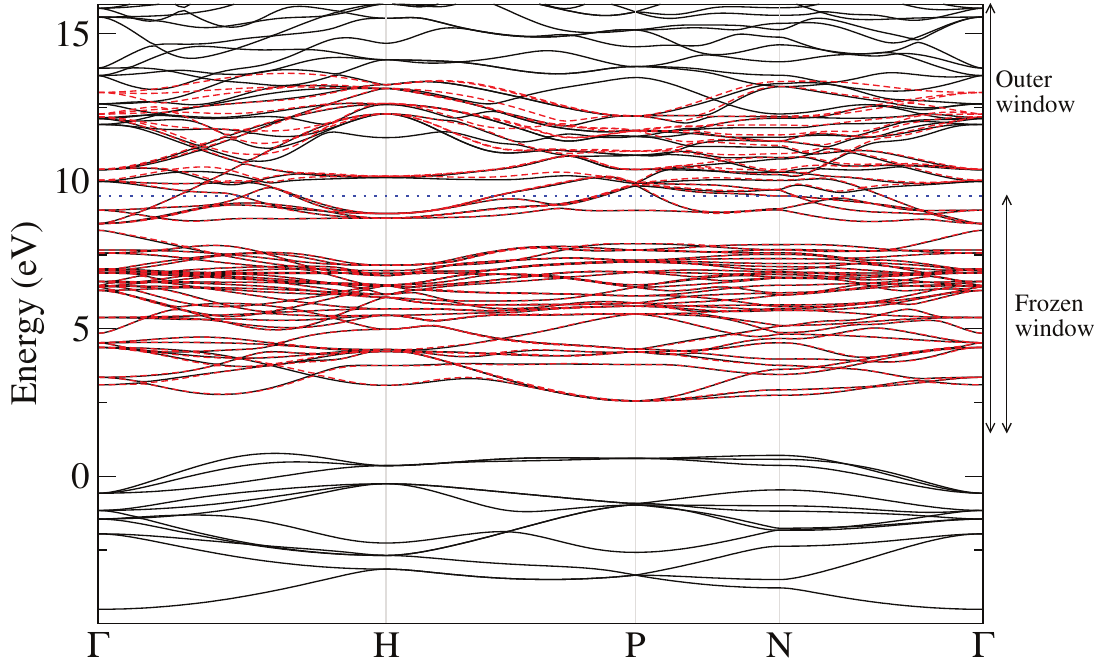}
\caption{Comparison between the band structure of CoSb$_3$ obtained from first-principles (black lines) and the Wannier functions interpolation with 56 WFs, as described in the text (red dashed lines). The top of the frozen energy window is indicated by the blue dotted line. The outer and the inner (frozen) windows are also indicated. The lowest 12-band manifold is irrelevant for transport purposes and is excluded from the WF calculation.
\label{fig:CoSb3-band-comparison}}
\end{figure}

MLWFs are calculated using the \textsc{Wannier90} code~\cite{Mostofi:2008}.
As can be seen in Fig.~\ref{fig:CoSb3-band-comparison}, where the band structure of CoSb$_3$ is plotted with black lines along the $\Gamma$--H--P--N--$\Gamma$ path, CoSb$_3$ is a semiconductor with two isolated valence manifolds of 12 and 36 bands respectively, and a conduction manifold separated from the valence bands by a (LDA) band gap of $\approx 0.2$~eV.
If one is interested in disentangling the valence band only, one can obtain 48 MLWFs which can be grouped in 12 Co states of $t_{2g}$ symmetry, 12 Sb--Sb bonding states and 24 Co--Sb bonding states, as discussed in Ref.~\cite{Volja:2012}.
However, we are here interested in obtaining WFs that are able to reproduce also the lowest-energy conduction states and we must then use the disentanglement procedure discussed in Sec.~\ref{sec:wf-basis}.

To simplify the problem, we first observe that only bands within a few $KT$ from the Fermi level are relevant for transport properties, due to the factor $\left(-\frac{\partial f(E,\mu,T)}{\partial E}\right)$ in Eqs.~\eqref{eq:eq-sigma}, \eqref{eq:eq-seebeck} and \eqref{eq:eq-thermal}.
Therefore the low-lying manifold of 12 valence states (which are primarily due to Sb $s$ states) can be safely disregarded for our aim, and we exclude the corresponding 12 bands when calculating the WFs.
We then choose the frozen window energy $E_{\text{froz}}$ approximately 1--1.5 eV above the bottom of the conduction manifold, which is sufficient for our transport purposes, and as initial guess for the WFs we employ atom-centered Gaussian-type orbitals: 5 $d$ states on each of the four Co atoms of the unit cell, and 3 $p$ states on each of the twelve Sb atoms (i.e., a total of 56 WFs).
With this set of parameters, we calculate the MLWFs using the iterative minimization algorithm of \textsc{Wannier90}. This choice for the initial guess and for $E_{\text{froz}}$ turns out to be able to provide real-valued and well-localized WFs, whose final spread lies within $0.6-1.0$~\AA$^2$ for $d-$type orbitals and within $3.4-3.8$~\AA$^2$ for $p$-type orbitals.

\begin{figure}[tbp]
\centering
  \subfigure[$d$-type Wannier function centered on a Co atom.]{\includegraphics[width=0.45\textwidth]{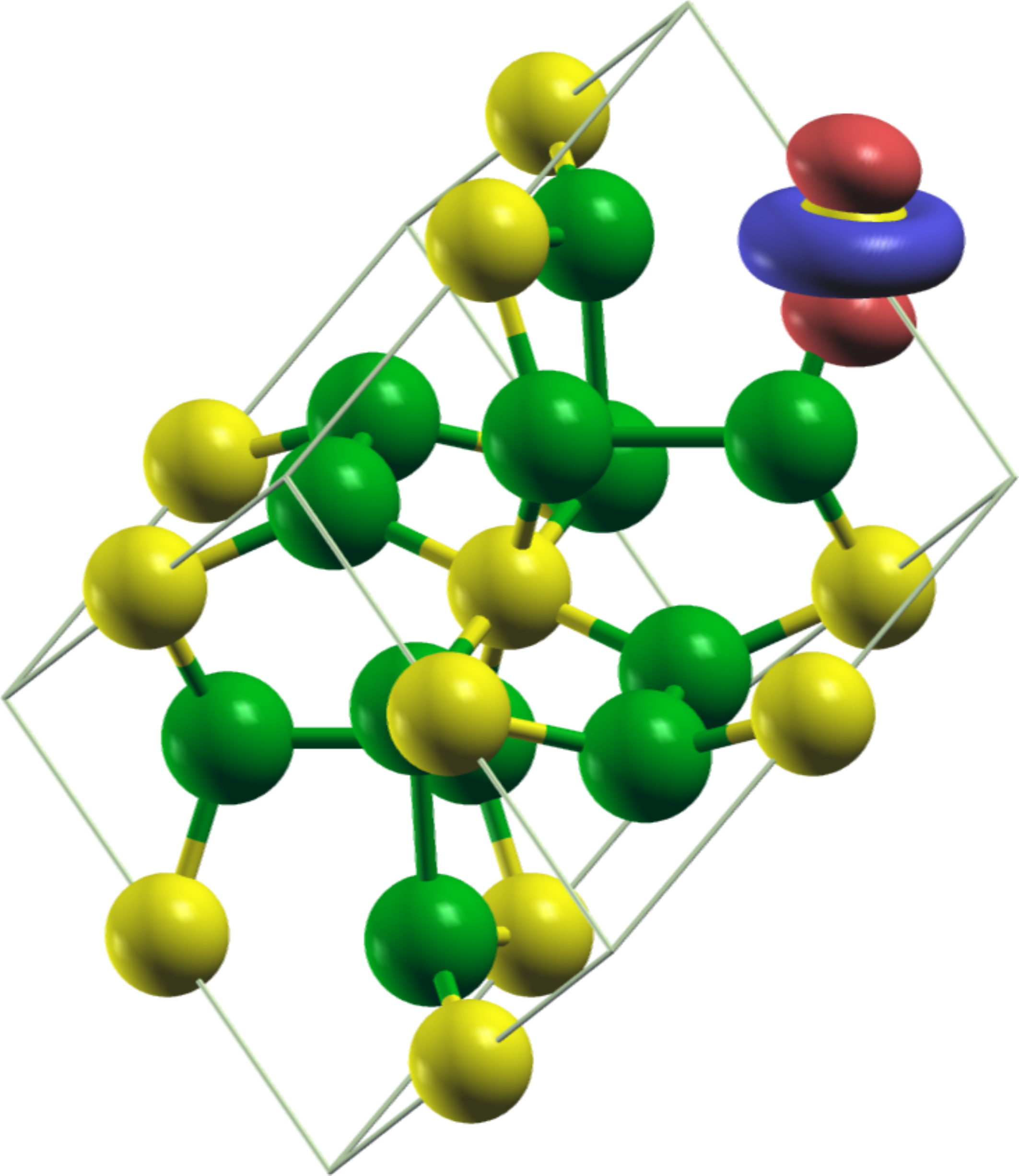}}\quad
  \subfigure[$p$-type Wannier function centered on a Sb atom.]{\includegraphics[width=0.45\textwidth]{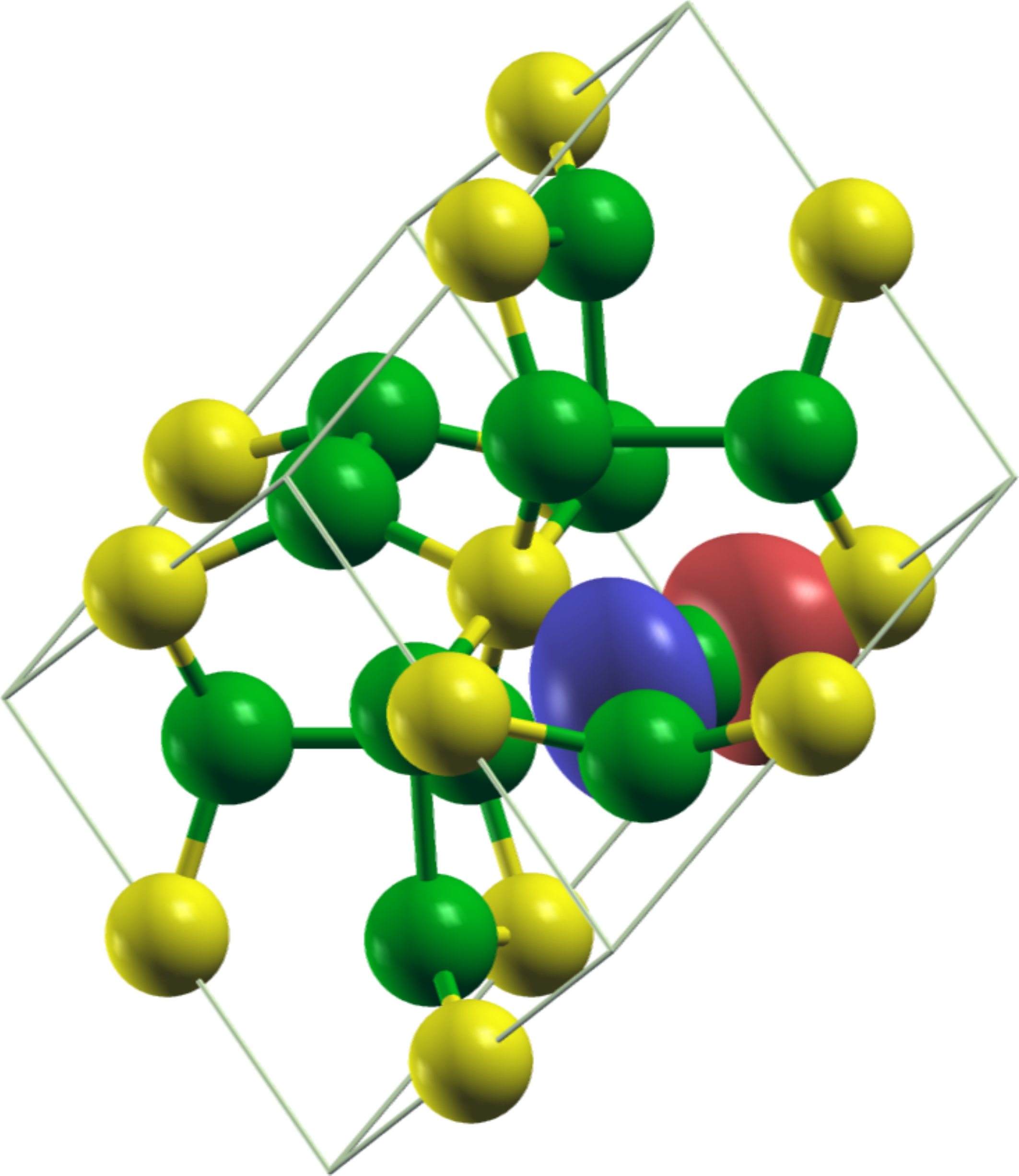}}
\caption{Real-space plot of MLWFs in CoSb$_3$. Co atoms are represented in yellow, Sb atoms in green.  \label{fig:CoSb3-realspace-WF} The plots have been realized using XCrySDen~\cite{Kokalj:2003}.}
\end{figure}

In order to assess the MLWFs so obtained, we show in Fig.~\ref{fig:CoSb3-band-comparison} a comparison between the first-principles band structure (black lines) and the Wannier-interpolated band structure (red dashed lines). The energy windows chosen are also reported in the figure. Within the frozen window, the WF interpolation is indeed in excellent agreement with the first-principles results.
The final WFs retain to a large amount the initial symmetry and center positions, as can be seen by inspection. In particular, we report in Fig.~\ref{fig:CoSb3-realspace-WF} the real-space plot of the isosurfaces of two selected WFs (a $d-$type WF centered on a Co atom and a $p-$type WF centered on a Sb atom). 

\begin{figure}[tbp]
\centering
\includegraphics[width=10cm]{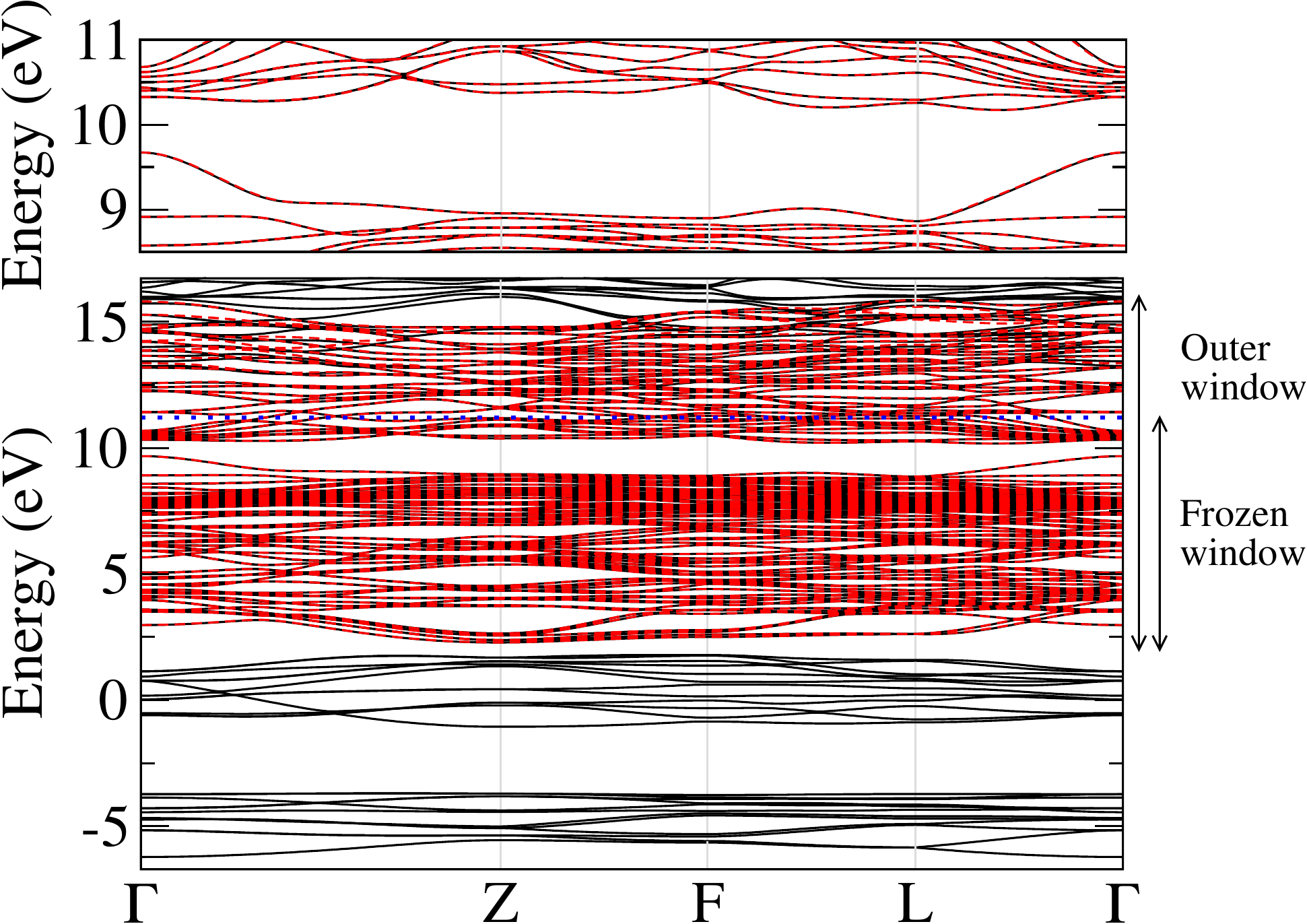}
\caption{Comparison between the band structure of CoGe$_{3/2}$S$_{3/2}$ obtained with the first-principles code (black lines) and the Wannier functions interpolation with 112 WFs, as described in the text (red dashed lines). The top panel displays the same bands, zoomed around the electronic gap. The top of the frozen energy window is indicated with a blue dotted line. The outer and the inner (frozen) windows are also indicated. The two lowest band manifolds are irrelevant for transport purposes and are excluded from the WF calculation. 
\label{fig:CoGeS-band-comparison}}
\end{figure}

For CoGe$_{3/2}$S$_{3/2}$ (whose cell contains twice as many atoms as the CoSb$_3$ cell) we follow a similar procedure in order to obtain MLWFs. In particular, we exclude the two lowest-lying valence manifolds, composed of 12+12 bands (see band structure in Fig.~\ref{fig:CoGeS-band-comparison}), and we extract 112 WFs from $d$ orbitals centered on Co atoms and $p$ orbitals centered on Ge and S atoms. Also in this case the top of the frozen window is set 1~eV above the bottom of the conduction band. 
The final spread of these states is within $2.1-4.4$~\AA$^2$; the comparison of first-principles and interpolated band structures is shown in Fig.~\ref{fig:CoGeS-band-comparison}.

\subsection{Code results and verification}
\label{sec:results-and-verification}
Once the MLWFs are obtained,  we can use them as a basis set to interpolate bands and band velocities as described in Sec.~\ref{sec:wf-basis} and calculate transport properties using the \textsc{BoltzWann} module introduced here.
Starting from the  $4 \times 4 \times 4$ coarse grid used for the construction of the Wannier functions, we interpolate the bands on a dense $40 \times 40 \times 40 $ mesh and calculate the TDF $\bm \Sigma(E)$ on this mesh. To sample $\bm \Sigma(E)$, we use a bin width of 1~meV. Moreover, for both material systems, we choose a relaxation time $\tau=10$~fs.

As a benchmark of the code, we compare our results for both systems with those obtained with the \textsc{BoltzTraP} code~\cite{Madsen:2006}. \textsc{BoltzTraP} is based on a smoothed Fourier interpolation of the bands~\cite{Pickett:1988}, and finite differences are used to evaluate band derivatives. While this may lead to incorrect results for the derivatives at band crossings, it has been shown on selected systems to be negligible if the $\vec k$ sampling is dense enough~\cite{Madsen:2006}.
In the \textsc{BoltzTraP} calculations, we start a $20 \times 20 \times 20$ $\vec k$ mesh from a non-self-consistent calculation performed with \textsc{Quantum ESPRESSO}. 

\begin{figure}[p]
\centering
\subfigure[$\sigma$ of CoSb$_3$\label{fig:CoSb3-elcond}]{
 \includegraphics[width=0.44\textwidth]{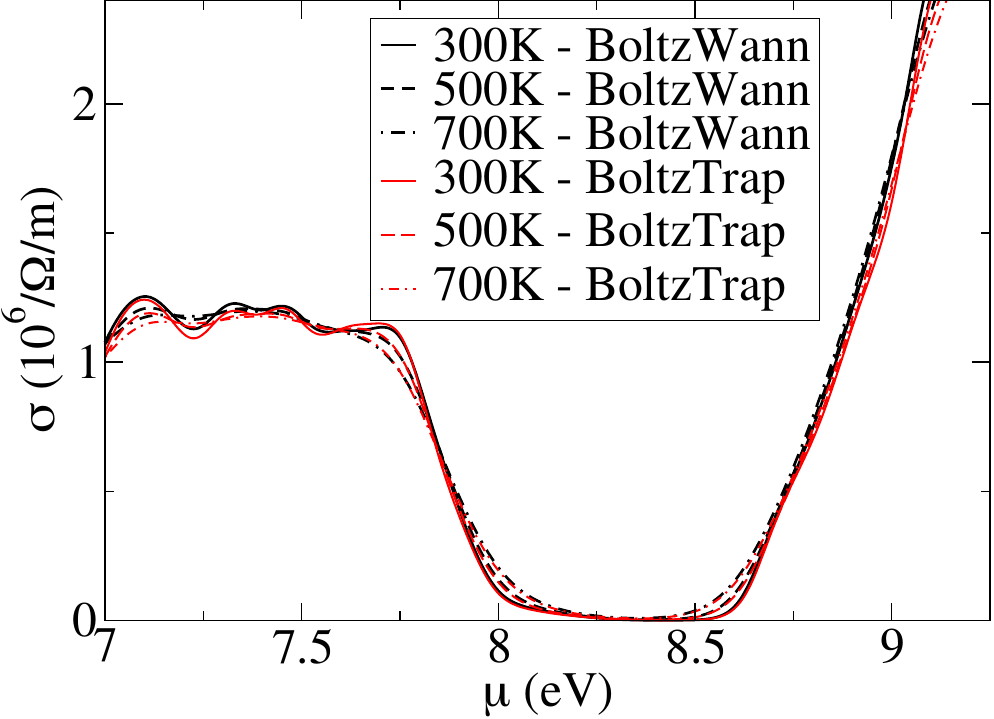}
 }
\subfigure[$\sigma$ of CoGe$_{3/2}$S$_{3/2}$\label{fig:CoGeS-elcond}]{
 \includegraphics[width=0.44\textwidth]{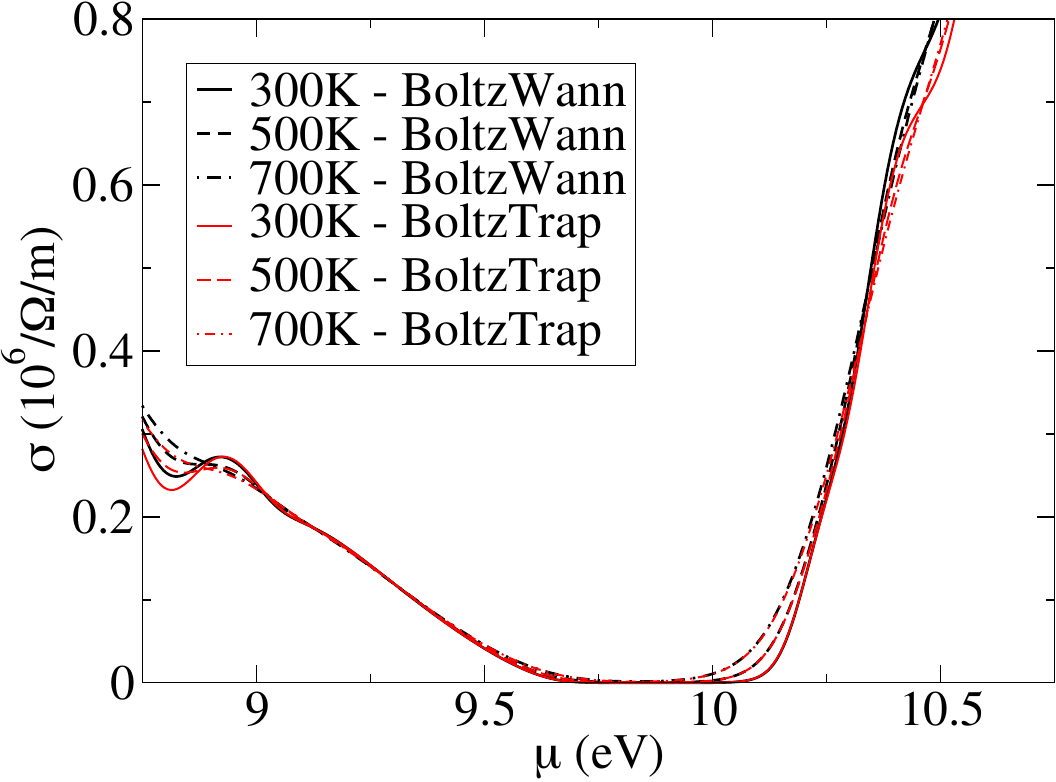}
 }
\caption{Comparison between the electrical conductivity $\sigma$ calculated with our \textsc{BoltzWann} code (black lines) and the \textsc{BoltzTraP} code~\cite{Madsen:2006} (red lines). The calculations have been performed at the three temperatures of 300, 500 and 700~K.}
\end{figure}

\begin{figure}[p]
\centering
\subfigure[$S$ of CoSb$_3$\label{fig:CoSb3-seebeck}]{
 \includegraphics[width=0.44\textwidth]{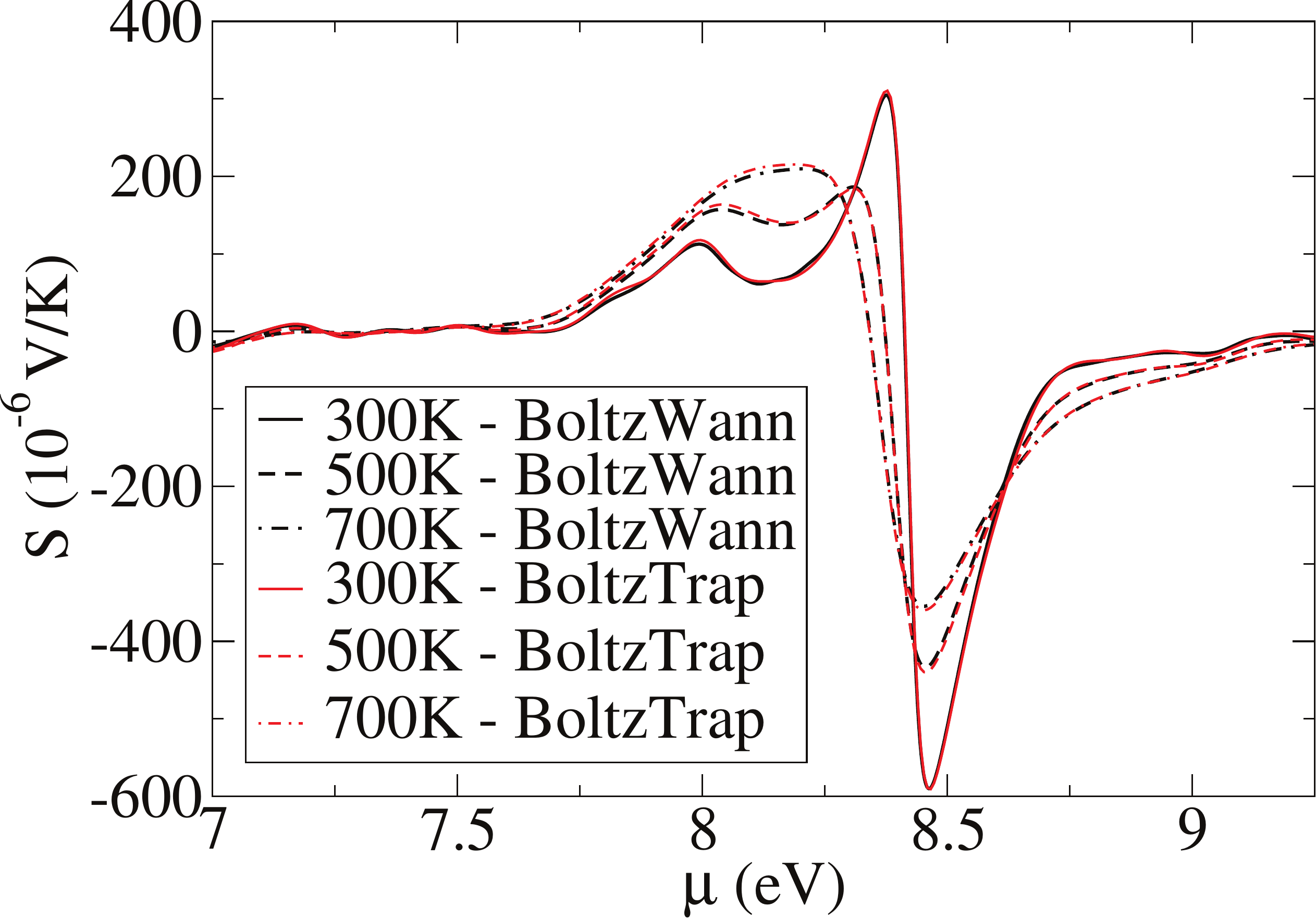}
 }
\subfigure[$S$ of CoGe$_{3/2}$S$_{3/2}$\label{fig:CoGeS-seebeck}]{
 \includegraphics[width=0.44\textwidth]{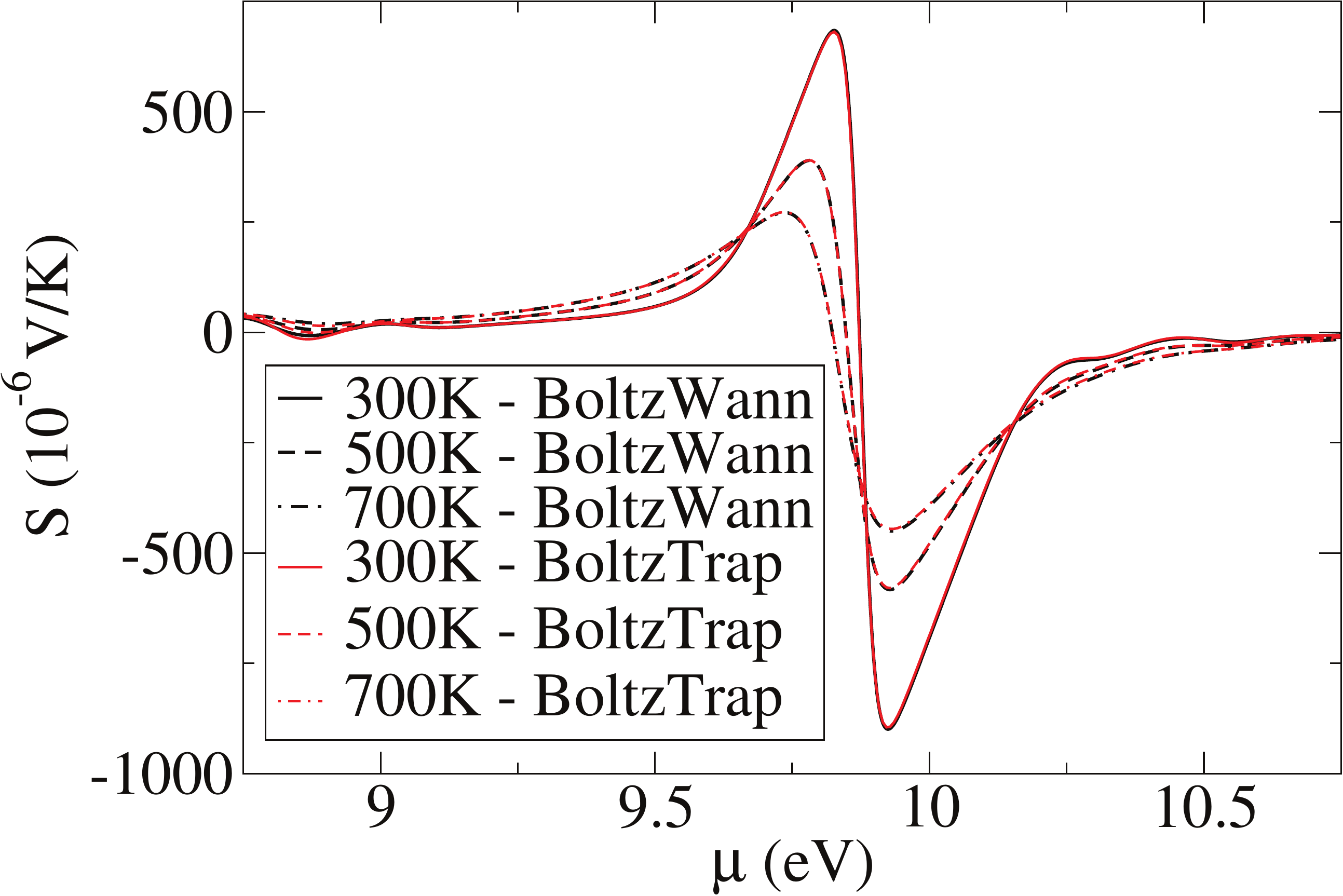}
 }
\caption{Comparison between the Seebeck coefficient $S$ calculated with our \textsc{BoltzWann} code (black lines) and the \textsc{BoltzTraP} code~\cite{Madsen:2006} (red lines). The calculations have been performed at the three temperatures of 300, 500 and 700~K.}
\end{figure}

\begin{figure}[p]
\centering
\subfigure[$K$ of CoSb$_3$\label{fig:CoSb3-thermcond}]{
 \includegraphics[width=0.44\textwidth]{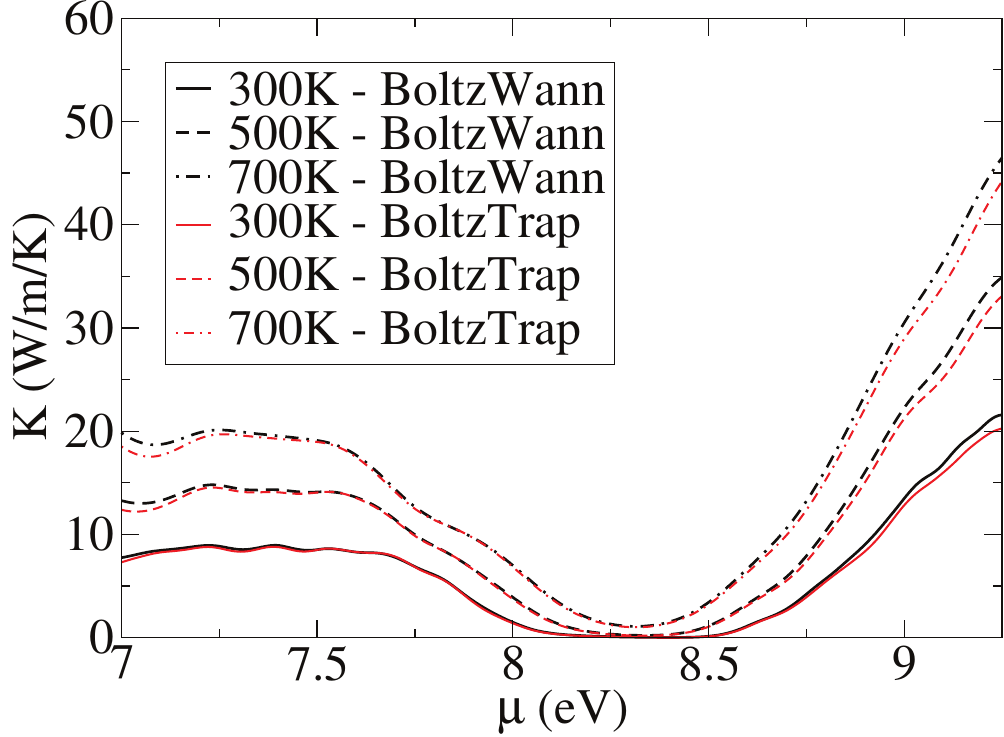}
 }
\subfigure[$K$ of CoGe$_{3/2}$S$_{3/2}$\label{fig:CoGeS-thermcond}]{
 \includegraphics[width=0.44\textwidth]{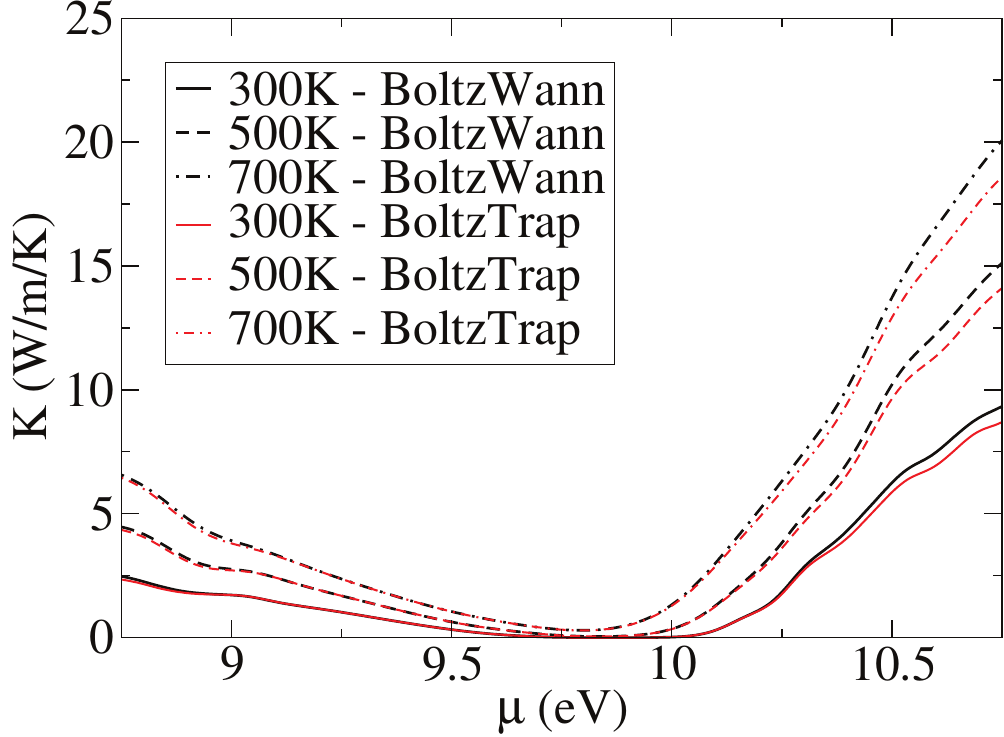}
 }
\caption{Comparison between the coefficient $K$ of Eq.~\eqref{heatcurrent} calculated with our \textsc{BoltzWann} code (black lines) and the \textsc{BoltzTraP} code~\cite{Madsen:2006} (red lines). The calculations have been performed at the three temperatures of 300, 500 and 700~K.}
\end{figure}

In Fig.~\ref{fig:CoSb3-elcond} we plot the electrical conductivity for CoSb$_3$ as a function of the chemical potential $\mu$ for the three temperatures $T=300$, $500$ and $700$~K, and compare these results with those obtained by \textsc{BoltzTraP} code, showing excellent agreement. Analogous plots for the Seebeck coefficient $S$ and the $K$ coefficient of Eqs.~\eqref{heatcurrent} and \eqref{eq:eq-thermal} are reported in Fig.~\ref{fig:CoSb3-seebeck} and Fig.~\ref{fig:CoSb3-thermcond}, respectively.
The results for CoGe$_{3/2}$S$_{3/2}$ are shown in Fig.~\ref{fig:CoGeS-elcond}, Fig.~\ref{fig:CoGeS-seebeck} and Fig.~\ref{fig:CoGeS-thermcond} for electrical conductivity, Seebeck coefficient and $K$ coefficient, respectively. Also in this case the agreement between the two codes is very good, providing a validation of \textsc{BoltzWann}. 
We also emphasize here that, since the first-principles band structure is reproduced correctly by the Wannier interpolation only within the frozen window, discrepancies between the results of the two codes are to be expected when $\mu$ is outside the frozen window.

\begin{figure}[tbp]
\centering \includegraphics[width=10cm]{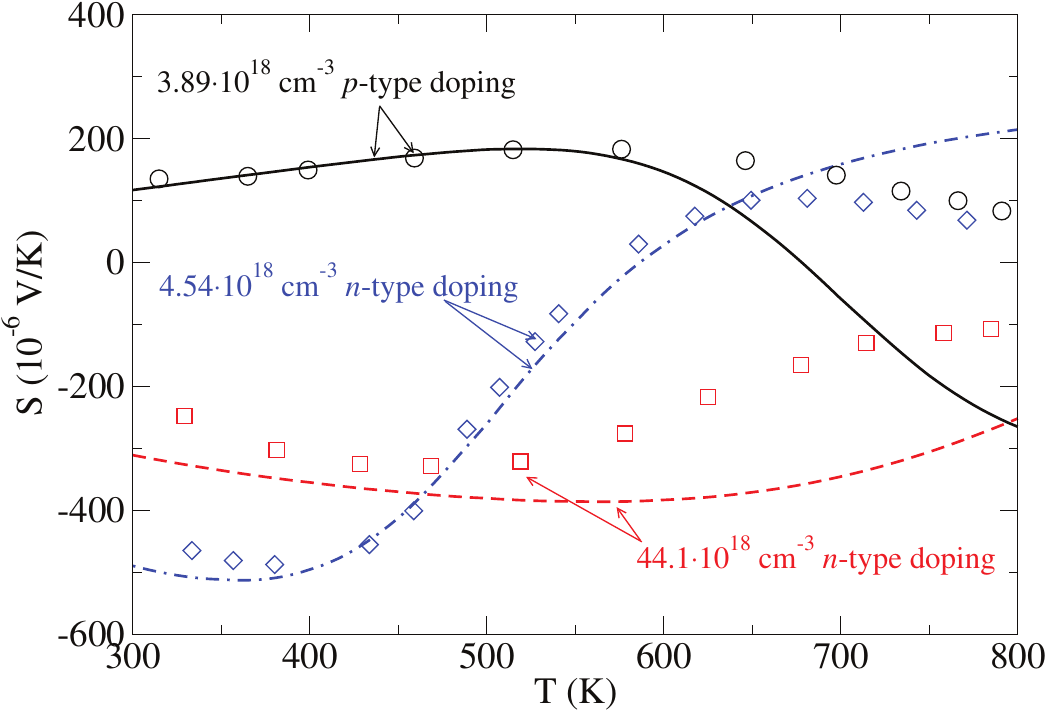}
\caption{Comparison between the Seebeck coefficient calculated with our \textsc{BoltzWann} code (lines) with the experimental results from Ref.~\cite{Caillat:1996} for $n$- and $p$-doped CoSb$_3$. The experimental data correspond to samples 2NB13 (3.89$\cdot 10^{18}$ cm$^{-3}$ $p$-type doping, black circles), 1CS10 (4.54$\cdot 10^{18}$ cm$^{-3}$ $n$-type doping, blue diamonds) and 4OB25 (44.1$\cdot 10^{18}$ cm$^{-3}$ $n$-type doping, red squares) of Ref.~\cite{Caillat:1996}. Theoretical results are obtained assuming the experimental doping value and excluding effects due to intrinsic carriers.
\label{fig:exp-comparison-CoSb3}}
\end{figure}

It is also instructive to compare our results with some experimental measurements, despite the known limitations of Boltzmann transport theory~\cite{Allen:1996}, and of LDA in describing correctly the band gap of semiconductors. 
In Fig.~\ref{fig:exp-comparison-CoSb3} we compare in particular the experimental results for two (Te or Pd) $n$-doped CoSb$_3$ samples and one as-grown $p$-doped sample from Caillat \emph{et al.}~\cite{Caillat:1996}. 
In order to reproduce the experimental results for the Seebeck coefficient as a function of the temperature, we have to find consistently the chemical potential $\mu(T)$ that reproduces the experimental doping level. To this aim, we use the \textsc{BoltzWann} code to calculate the Seebeck coefficient $S(\mu_i,T_j)$ on a grid of $\mu_i$ and $T_j$ values, and we also calculate at the same time the density of states (DOS). For each given temperature $T$, we then integrate the DOS times the Fermi distribution function and we find the value of the chemical potential $\mu(T)$ that reproduces the experimental doping by means of a bisection algorithm. In this calculation, we neglect the effects of minority carriers. Finally, we interpolate the $S(\mu_i,T_j)$ grid obtained with \textsc{BoltzWann} to obtain the value $S(\mu(T),T)$ that is plotted in Fig.~\ref{fig:exp-comparison-CoSb3}.
We observe a very good agreement at low temperatures, as also demonstrated before by Wee {\it et al.}~\cite{Wee:2010} for one of the experimental samples. At higher temperatures, deviations from the experiment are expected and can be at least partially attributed to the minor carriers starting to play a significant role~\cite{Sharp:1995}. Moreover, while in the constant relaxation-time approximation the Seebeck coefficient $S$ does not depend on $\tau$, actual variations of $\tau$ over the Brillouin zone may also be a cause of discrepancy.

\section{Conclusions}
We have implemented and tested a new Fortran module to obtain transport properties (electrical conductivity, Seebeck coefficient, electronic thermal conductivity) in a semiclassical transport framework and using a maximally-localized Wannier function basis to interpolate band structures and band velocities.
We have verified the results of our code on the two skutterudite systems CoSb$_3$ and CoGe$_{3/2}$S$_{3/2}$ comparing the results with the publicly available \textsc{BoltzTraP} code, and
have found a very good agreement of the results obtained with the two codes.
A major advantage of this approach is the increased accuracy of the results due to the proper treatment of band crossings, thanks to the possibility of obtaining analytical expressions for the band derivatives in the Wannier functions basis, and a much reduced computational cost. These advantages become even more relevant or essential in the case of very large systems, where many intersecting folded bands in the Brillouin zone can be found. Moreover, provided a model for the dependence of $\tau$ on the band index $n$ and on the quasimomentum $\vec k$, the code can also be easily extended beyond the constant relaxation-time approximation.
The code has been included as a \textsc{BoltzWann} module inside the existing \textsc{Wannier90} code. It can therefore be used in combination with any code that provides an interface to \textsc{Wannier90} such as Quantum ESPRESSO, ABINIT, SIESTA, FLEUR, WIEN2k, VASP, CASTEP, \ldots 

The authors gratefully acknowledge A. Cepellotti, G. Chen, S. Halilov, A.~A. Mostofi,  X. Qian, Z.~F. Ren, I. Souza, D. Wee, and J.~R. Yates for valuable discussions. 
This work was carried out as part of the MIT Energy Initiative, with financial support from Robert Bosch LLC.

\end{document}